\pdfoutput=1  
\documentclass[aps, prl, superscriptaddress, floatfix,onecolumn, notitlepage, preprint, nobibnotes, showkeys]{revtex4-2}
\usepackage{amsmath,amssymb,amsfonts}
\usepackage{hyperref}
\usepackage{graphicx}
\usepackage{lipsum}
\usepackage{ulem}
\usepackage[table,x11names]{xcolor}
\usepackage{gensymb}
\usepackage{bm}
\usepackage{physics}
\usepackage{colortbl}
\usepackage{booktabs}
\usepackage{rotating}

\begin{document}
\def\figSchematic{
    \begin{figure} [h]
    	\includegraphics[height=2.75 in,angle=0]{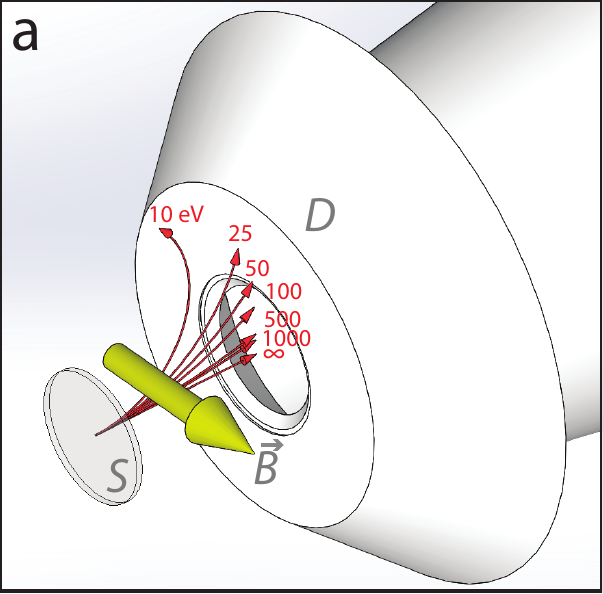}
        \includegraphics[height=2.75 in,angle=0]{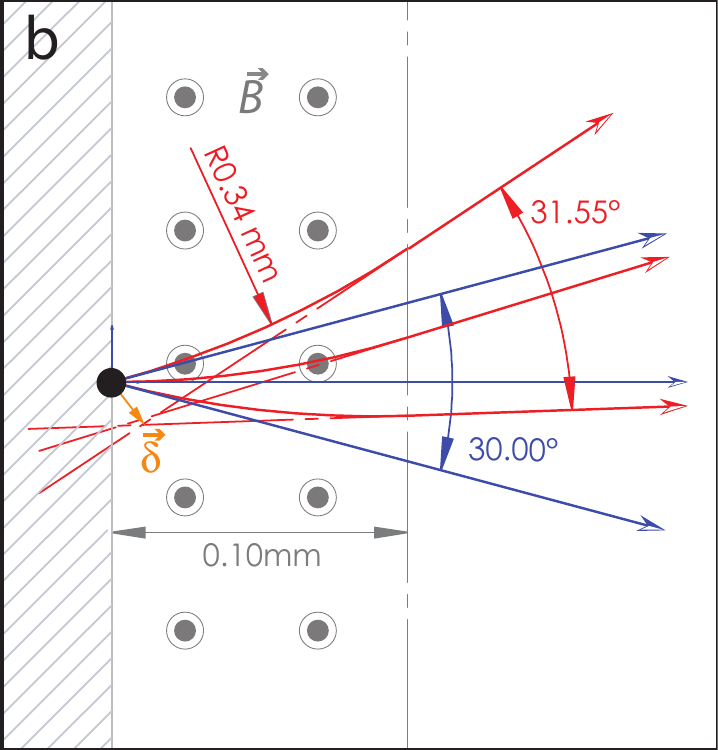}
     
    	\caption{\textbf{Schematic of MagnetoARPES Experiment.} [color in print] (a) Electron paths for photoelectrons emitted at normal angle to the surface from a sample ($S$) in a uniform magnetic field \vecB\, = 0.5 mT. The input lens element of an electron detector ($D$) is at working distance of 35 mm.  Electron trajectories are calculated for various kinetic energies as indicated. (b) Electron paths for photoelectrons emitted at 0\degr\ and $\pm$ 15\degr\ from normal for KE = 100 eV and a uniform magnetic field  confined to a distance 100 \um\ from the sample surface.  Red (blue) rays indicate \absB\, = 100 mT (\absB\, = 0).  The perturbed rays projected back to a virtual source are indicated by dashed lines.  These back-projected rays converge approximately to a virtual source located \vecd\, from the original emission source of source size 10 \um\ indicated by the black dot.
    	\label{fig:figSchematic}}
    \end{figure}
}
\def\figSchematicRef{Fig.\ \ref{fig:figSchematic}}

\def\figDevice{
    \begin{figure*} [tb]
    \includegraphics[height=2.75in]{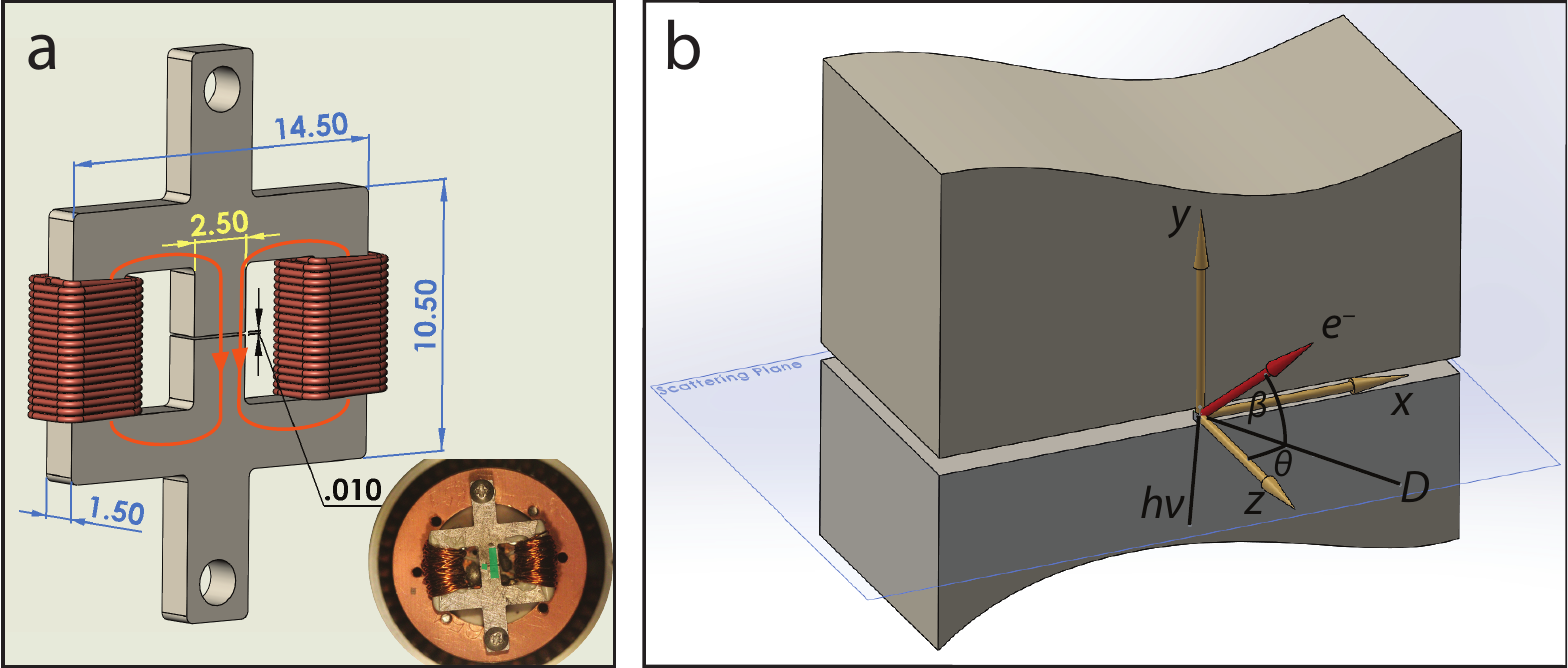}
        \caption{\textbf{Magnetic Device Constructed from Nickel Plate.} [color in print] (a) schematic of typical assembled magnetic device. Typically 60 turns of wire are wrapped on each arm.  Coils are energized so that magnetic fluxes add as indicated (field direction \bdown\ is indicated). The inset shows a similar device mounted onto a 25.4 mm sample carrier.  The central gap is 100 \um\ wide; samples are affixed over the central gap by silver epoxy. Samples are graphene/SiC(0001) false-colored green in this example.  All dimensions are in mm.  (b) Magnified view of yoke gap, with coordinate system and photoemission geometry.  The photon ray $h\nu$ and detector central axis $D$ are fixed and lie in the blue scattering plane. The device is rotated with polar angle $\theta$ about the $y$-axis and can be positioned arbitrarily in $(x, y, z)$. The detector simultaneous collects electrons with polar angle distribution $\pm$ 15\degr and using internal deflection plates, can select electrons emitted with elevation angle $\beta$ in the range $\pm$ 15\degr.
        \label{fig:figDevice}}
    \end{figure*}
}
\def\figDeviceRef{Fig.\ \ref{fig:figDevice}}

\def\figMag{
    \begin{figure*}[tb]
    \includegraphics[width=6.25in]{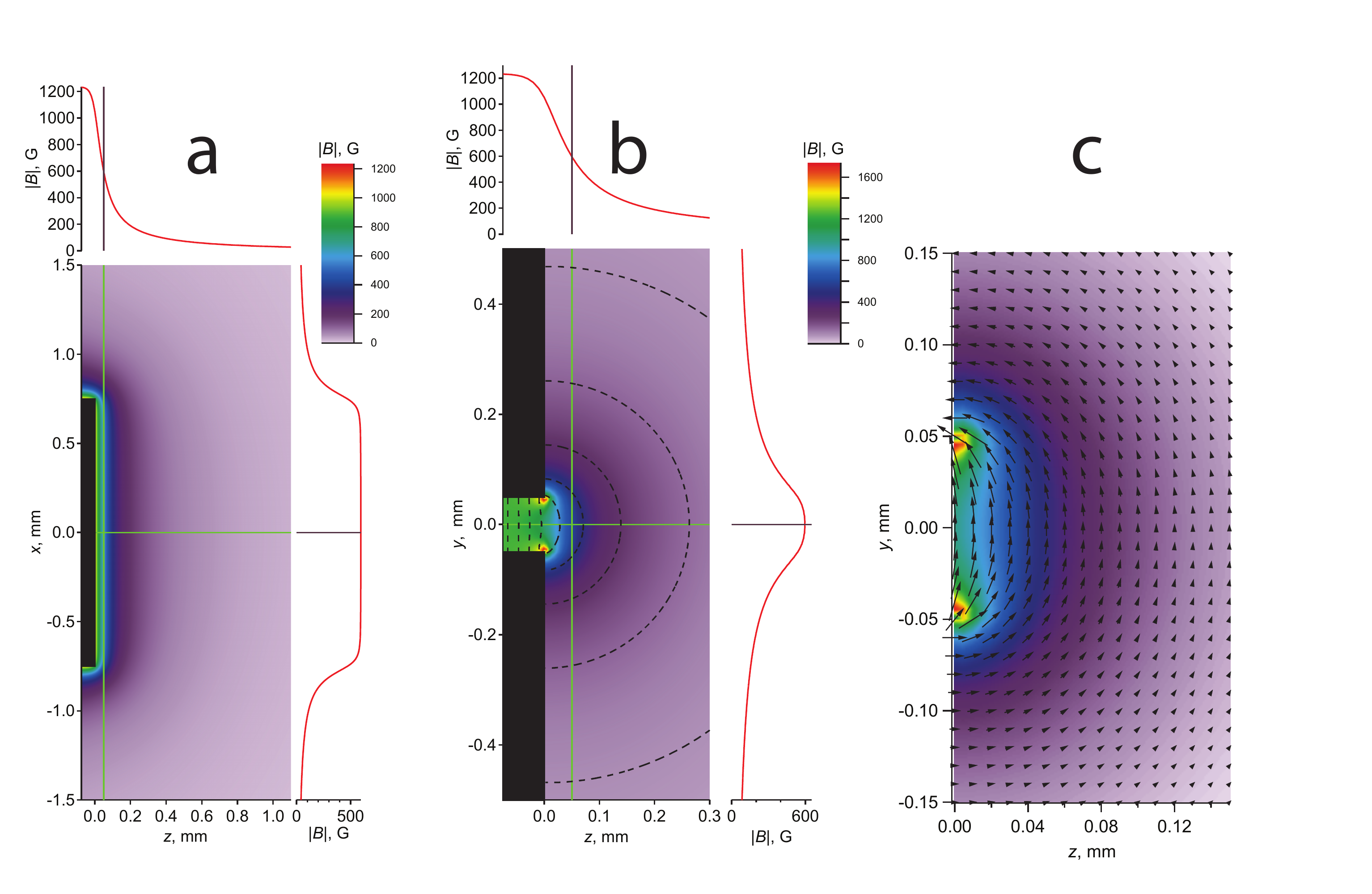}
        \caption{\textbf{Calculated Magnetic Field Patterns.} [color in print] (a, b) total magnetic field calculations within the planes (a) horizontal plane $y=0$ and (b) vertical plane $x=0$, resp.\ The red profiles are  the magnetic field strength along lines indicated by the green lines overlaying the magnetic field maps. The green line along $z=50$ \um\ represents a typical sample surface height.  The black rectangles in (a, b) indicate the projected positions of the yoke body.  The dashed lines in (b) indicate the magnetic flux lines.  Calculations are for \bup=\borigin\ = \boriginG\ at $(x,y,z)=(0,0,0)$.  (c) a magnified view of the data in (b), including a vector field map indicating strength and direction of \vecB. Blue dashed lines indicate the position of the magnetic gap.
        \label{fig:figMag}}
    \end{figure*}
}
\def\figMagRef{Fig.\ \ref{fig:figMag}}

\def\figDistort{
    \begin{figure*}[tb]
    \includegraphics[width=6.25in]{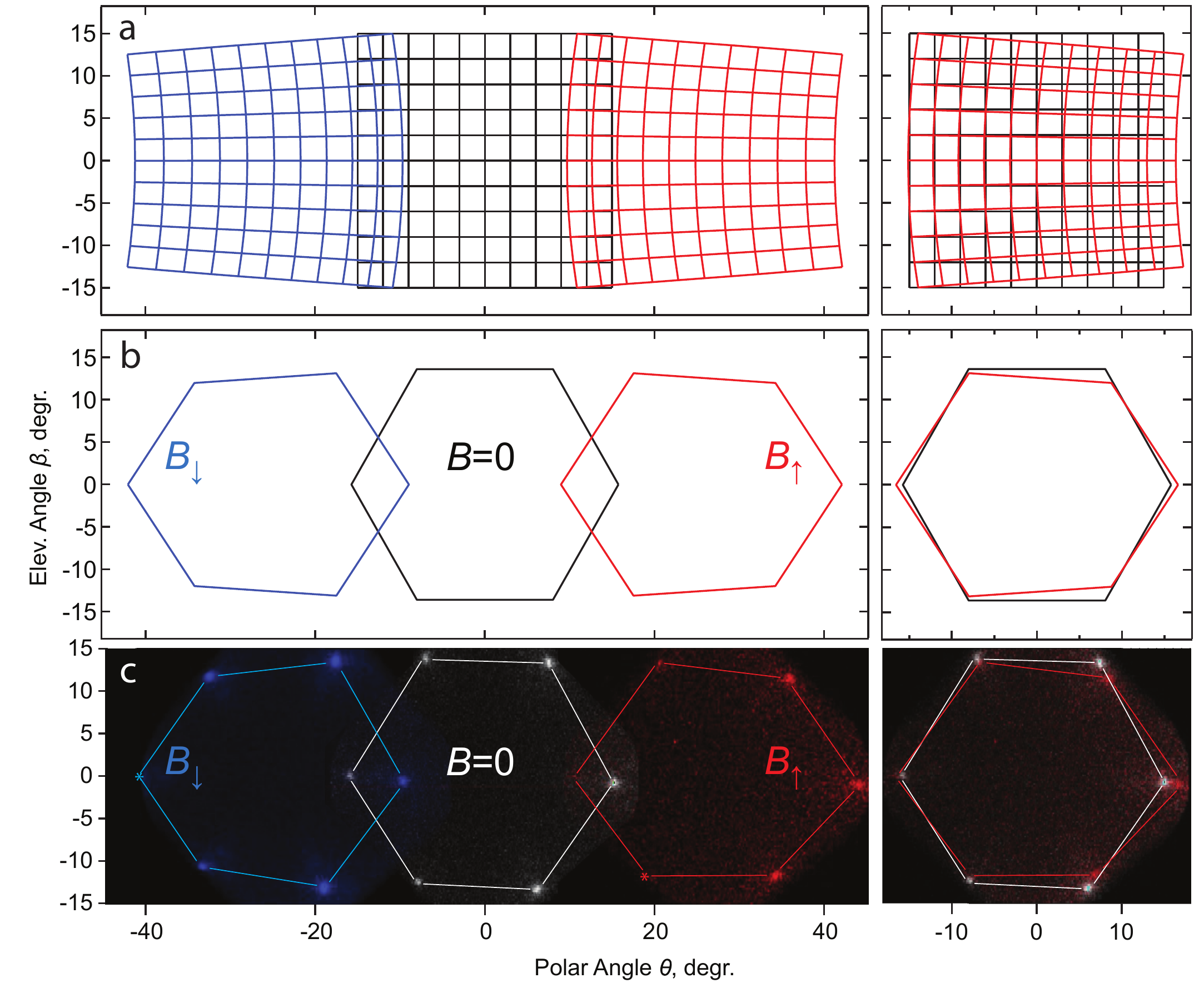}
        \caption{\textbf{MagnetoARPES Angle Distribution Maps} [color in print] for (a) a uniform set of electron rays calculated for a 3\degr $\times$ 3\degr mesh, (b) the graphene Dirac points (calculated), and (c) the graphene Dirac points (experimental).  The unperturbed trajectories are indicated by black or white; the trajectories with downwards field \bdown\ are indicated in blue; the trajectories in upwards field \bup\ are indicated in red.  Note that vertices  are calculated/observed; the connecting lines are guides to the eye.  The left panels show the trajectories in absolute angle space, while the right panels overlay \bup\ and $B=0$ trajectories for comparison.  Calculation details: Magnetic field was \borigin\ at the surface of the magnetic yoke, and \bsample\ at the surface of the sample, which was 50 \um\ thick. Kinetic energy was 150 eV.   Experimental details: \hv=150 eV, shown are energy-integrated angle distributions for a 0.15meV window around the graphene Dirac energy.  Experiments for \bup, \bdown, and $B=0$ were conducted separately after separately centering electrons at \G\ using the manipulator stepping motor to adjust $\theta$.  Resulting angle distributions are overlayed with spacing according to the stepping motor positions. Points marked ``*'' are outside the detector acceptance but their positions are inferred from data at higher binding energy.
        \label{fig:figDistort}}
    \end{figure*}
}
\def\figDistortRef{Fig.\ \ref{fig:figDistort}}

\def\figMagYZ{
    \begin{figure*}[tb]
    \includegraphics[width=6.25in]{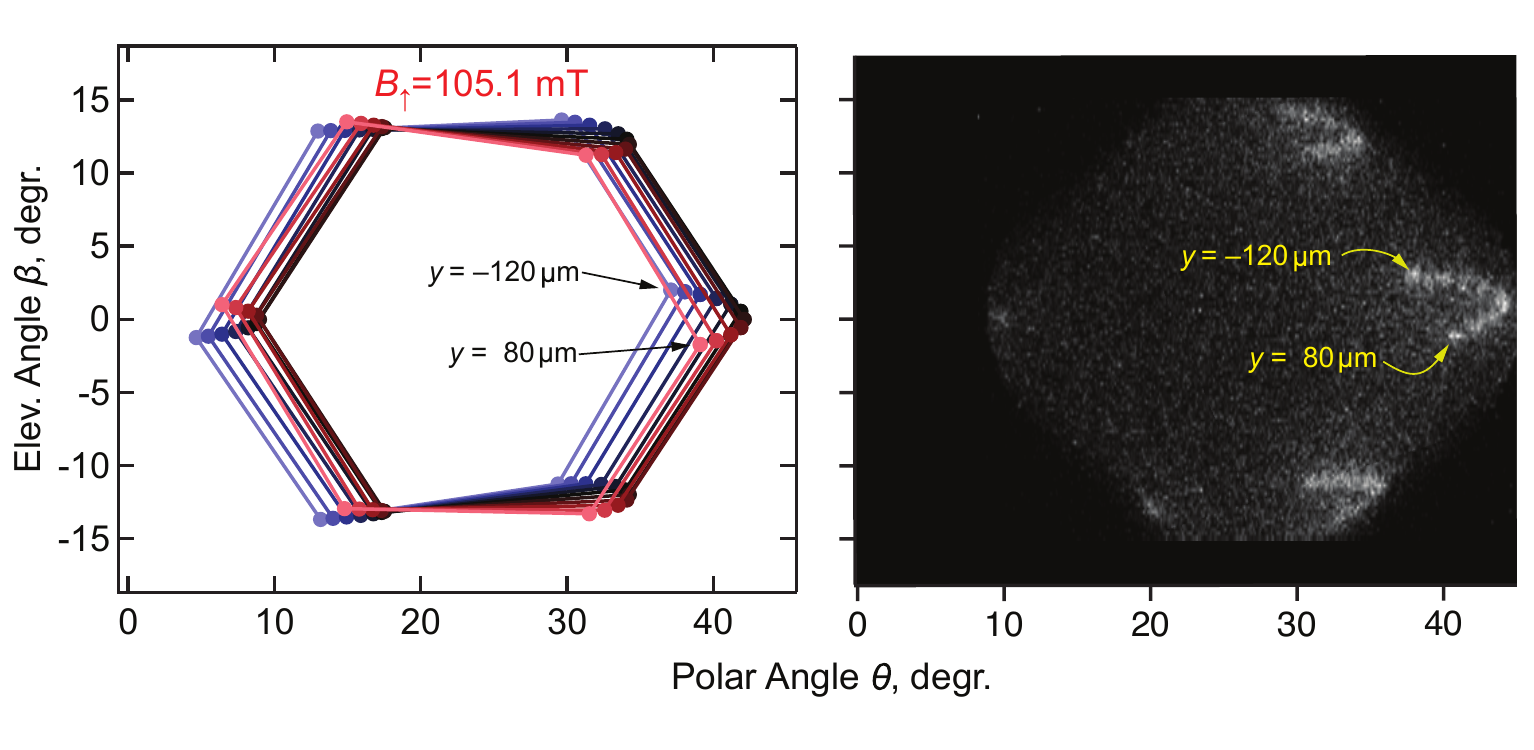}
        \caption{\textbf{Effect of Out--of-Plane Magnetic Field Direction.} [color in print] (a) Calculated K-points of graphene as a function of $y$ measured in 20 \um\ increments relative to the center of the 100 \um\ gap for \hv=150 eV.  Magnetic field configuration is \bup=\borigin\  at the origin, or $B_y=$\bsample\  at the sample surface at $(x, y, z)=$(0, 0, 50 \um).  Symbols reflect electron trajectories; lines are a guide to the eye only.  b) Experimental measurements of the graphene Dirac crossing points under the same conditions as (a).  Measurements of the ARPES angle distribution for each $y$ value are summed to create a composite map showing the locus of each K point position as a function of $y$.   
        \label{fig:figMagYZ}}
    \end{figure*}
}
\def\figMagYZRef{Fig.\ \ref{fig:figMagYZ}}

\def\figFocus{
    \begin{figure*}[tb]
    \includegraphics[height=3in]{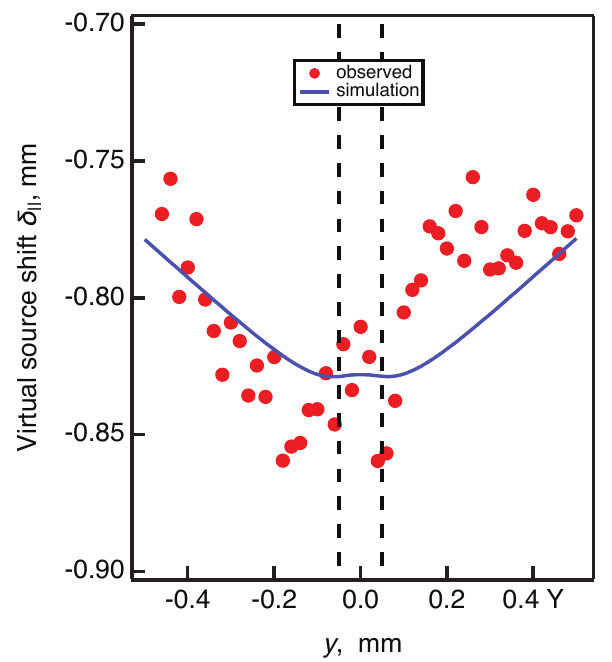}
        \caption{\textbf{Virtual Source Shift.} [color in print] The blue line indicates the predicted shift in virtual source position \dpar\ projected onto the surface plane as a function of $y$, where \bup= \borigin\ at $(x,y,z)=(0,0,0)$ and $B_y$=\bsample\ at the sample surface (0,0,50 \um).  Simulations are for electrons with kinetic energy KE=130 eV.    The red symbols are the experimentally observed values of \dpar, obtained using real-space (``transmission'') imaging mode of the analyzer, measured for electrons with kinetic energy KE=130 eV. Dashed lines indicate the edges of the yoke gap. 
        \label{fig:figFocus}}
    \end{figure*}
}
\def\figFocusRef{Fig.\ \ref{fig:figFocus}}

\def\eb{$E_\mathrm{B}$}  
\def\etal{\textit{et al.}}
\def\vs{\textit{vs}}
\def\aw{$A(\omega)$}
\def\akw{$A(\mathbf{k},\omega)$}
\def\akexpt{$A^{e}(\mathbf{k})$}
\def\awexpt{$A^{e}(\omega)$}
\def\akwexpt{$A^{e}(\mathbf{k},\omega)$}
\def\akwcalc{$A^{c}(\mathbf{k},\omega)$}
\def\skw{${\mathrm{\Sigma}(\mathrm{\mathbf{k}},\omega)}$}
\def\iskw{${\mathrm{Im\Sigma}(\mathrm{\mathbf{k}},\omega)}$}
\def\rskw{${\mathrm{Re\Sigma}(\mathrm{\mathbf{k}},\omega)}$}
\def\res{${\mathrm{Re\Sigma}}$}
\def\ims{${\mathrm{Im\Sigma}}$}
\def\wbk{$\omega_b(\mathbf{k})$}
\def\kvec{$\mathbf{k}$}
\def\degr{$^{\circ}$}
\def\um{$\mu$m}
\def\G{$\mathrm{\Gamma}$}
\def\hv{$h\nu$}
\def\uARPES{$\mathrm{\mu}$ARPES}

\def\kill #1{\sout{#1}}
\def\eli #1{\textcolor{red}{#1}}   
\def\elicomment #1{\eli{[ER: #1]}} 
\def\elireplace#1#2{\kill{#1}\eli{\ #2}}
\def\imskw{Im$\Sigma(\omega, k)$}


\def\vecB{$\vec{B}$}
\def\absB{$|\vec{B}|$}
\def\Bxyz{$\vec{B}(x,y,z)$}
\def\absBxyz{$|\vec{B}(x,y,z)|$}

\def\vecd{$\vec{\delta}$}
\def\dpar{$\delta_{||}$}
\def\dperp{$\delta_{\perp}$}
\def\bx{$B_x$}
\def\by{$B_y$}
\def\bz{$B_z$}
\def\ve{$\vec{v}_e$}
\def\bup{$B_\uparrow$}
\def\bdown{$B_\downarrow$}
\def\bupdown{$B_{\uparrow\downarrow}$}

\def\bpoles{123.5 mT}
\def\bpolesG{1235 G}
\def\borigin{105.1 mT}
\def\boriginG{1051 G}
\def\bsample{59.6 mT}
\def\als{Advanced Light Source, Lawrence Berkeley National Laboratory, Berkeley, CA 94720, USA}

\def\chemnitz{Center for Materials, Architectures and Integration of Nanomembranes (MAIN), Chemnitz University of Technology, 09126 Chemnitz, Germany}

\author{Sae Hee Ryu}
\affiliation{\als}

\author{Garett Reichenbach}
\affiliation{\als}
\affiliation{Department of Physics, University of Arizona, Tucson, AZ 85721, USA}

\author{Chris M. Jozwiak}
\affiliation{\als}

\author{Aaron Bostwick}
\affiliation{\als}

\author{Peter Richter}
\affiliation{\chemnitz}

\author{Thomas Seyller}
\affiliation{\chemnitz}

\author{Eli Rotenberg}
\affiliation{\als}

\date{\today}

\title{magnetoARPES: Angle Resolved Photoemission Spectroscopy with Magnetic Field Control} 

\maketitle

\textbf{Angle-Resolved Photoemission Spectroscopy (ARPES) is a premier technique for understanding the electronic excitations in conductive, crystalline matter, in which the induced photocurrent is collected and dispersed in energy and angle of emission to reveal the energy- and momentum-dependent single particle spectral function \akw. So far, ARPES in a magnetic field has been precluded due to the need to preserve the electron paths between the sample and detector.  In this paper we report progress towards ``magnetoARPES'', a variant of ARPES that can be conducted in a magnetic field. It is achieved by applying a microscopic probe beam ($\lesssim$  10 \um ) to a thinned sample mounted upon a special sample holder that generates magnetic field confined to a thin layer near the sample surface.  In this geometry we could produce ARPES in magnetic fields up to around $\pm$ 100 mT.  The magnetic fields can be varied from purely in-plane to nearly purely out-of-plane, by scanning the probe beam across different parts of the device.  We present experimental and simulated data for graphene to explore the aberrations induced by the magnetic field.  These results demonstrate the viability of the magnetoARPES technique for exploring symmetry breaking effects in weak magnetic fields.
}


\section{Introduction}
Angle resolved photoemission spectroscopy (ARPES) is a premier tool for understanding the excited states of crystalline materials.  It measures the single-particle spectral function \akw\ which encodes a material's occupied electronic bands, that are modified by many-body interactions. These interactions manifest in a finite spectral width associated with scattering, and in renormalized energies associated with virtual excitations in the surrounding medium . ARPES is furthermore very sensitive to the wavefunction symmetry, because the observed photocurrent is modulated by strong matrix element effects that depend on light polarization and orbital arrangement \cite{kevan_angle-resolved_1992, Damascelli2004, hufner_photoelectron_2007}.  ARPES is highly complementary to other probes of electronic properties such as transport and scanning tunneling microscopy (STM), because it provides momentum- (\kvec -) dependence of the obtained information. In addition, with the recent advent of small-probe ``nanoARPES'' endstations\cite{Bostwick2012,Rotenberg2014}, it is now possible to study real-space variations associated with inhomogenous materials as well as materials in device geometries.  Current state-of-the-art spatial resolution is in the range of $\sim 100$ - 1000 nm at synchrotrons\cite{Koch2018a}, and $\sim$ 3000 nm using laser-based ARPES setups\cite{cucchi_microfocus_2019}.  Several microARPES endstations capable of sub-10 \um\ ARPES using Kirkpatrick-Baez optics have also emerged in the last few years. 

Despite the popularity of ARPES for electronic structure determination, it is missing a key feature, which is the ability to measure \akw\ in magnetic fields.  Magnetism is deeply tied to symmetry, scattering,  and topology, and variations of the bandstructure with induced magnetic field can give very useful information. For example, it can suppress or enhance effects such as localization or superconductivity, break time-reversal symmetry,  and it can alter the topological class of a system.  We are presently motivated to study chiral anomaly affects in Weyl semimetals (WSMs).  These materials electronic structure in momentum space are characterized by pairs of Dirac-cone-like nodal points of opposite chirality.  In simultaneous crossed electric and magnetic fields, WSMs develop a chemical potential shift between the Dirac valleys, driving a chiral current in the material.  It is predicted that the chiral chemical potential, can reach on the order of 10 meV per 1 mT of applied magnetic field \cite{behrends_visualizing_2016}, depending on the material and its intervalley equilibration time.  Behrends \etal\ proposed that a confined fringing field at the surface of a gapped magnetic ``picture frame'' yoke could achieve usable fields in the mT regime, in which case the chiral currents could be visualized in ARPES according to the effect on the surface localized states.  Our aim for this study is to consider how a small probe beam could be used to increase the fields obtainable, to determine how the field can be measured and controlled, and to determine to what extent aberrations introduced by the field can be understood and controlled in a practical manner.

\figSchematic
 
In this paper, we show that modest magnetic fields can be applied to ARPES by taking advantage of the small spot sizes available at new nanoARPES and microARPES setups.  \figSchematicRef\  illustrates the basic principle for a setup employing one common electron analyzer with 35 mm working distance to the sample.  In \figSchematicRef(a) we illustrate why a uniform space-filling magnetic field makes ARPES impractical.  In a typical experiment, a photon beam strikes a sample ($S$), generating photoelectrons with a distribution of angles (from normal to grazing emission) and kinetic energies (KEs) from the workfunction cutoff up to the photon energy.  Suppose that the space between $S$ and the detector ($D$) entrance aperture is filled with a uniform magnetic field \absB\,= 0.5 mT = 5 G.  Electrons leaving the sample at normal emission angle towards the detector will be deflected upwards due to Lorentz force on the electrons.  \figSchematicRef(a) shows the resulting circular electron trajectories for kinetic energies in the range 10 - 1000 eV.  Of these, electrons below about 100 eV do not make it into the entrance aperture of the detector. Those electrons that are accepted will enter the entrance aperature at an extreme angle, where each value of kinetic energy  projects to a different source point far from the center axis of the detector input lens. The imaging of the electron trajectories by the lens onto the spectrometer entrance slit will be significantly aberrated in angle and energy, leading to poor resolution and sensitivity.  

\figSchematicRef(b) shows our proposed idea in an idealized geometry for typical ARPES conditions. Supposing instead of filling space, a uniform magnetic field is confined within  $\sim$ 100 \um\ layer above the sample surface.  We presume a 10 \um\ x-ray probe (black dot) is employed.  A typical hemispherical electron analyzer accepts a $\pm$ 15\degr\ electron distribution, indicated by the blue rays for the unperturbed situation \absB\,= 0.  In the case of \absB\,= 100 mT, and electron KE = 100 eV, these rays will be perturbed into circular paths (red rays) within the magnetic field region, beyond which the trajectories follow straight paths to the detector.   In this case the perturbed rays form a distribution very similar to the unperturbed rays, but with a rigid shift of the entire photoelectron distribution by about $15$\degr.  One only needs to rotate the detector relative to the sample in order to collect the magnetic-field perturbed ARPES signal, that we call the magnetoARPES spectrum.  At last two orders of magnitude higher field could be achieved compared to the space-filling field illustrated in \figSchematicRef(b).

A glance at \figSchematicRef(b) shows the kinds of aberrations that one may expect.  First, the greatest effect is the shift in angles corrected by a large rotation of the analyzer mentioned above. Second, the perturbed angle distribution is slightly wider (31.55\degr\ \vs\ 30.00\degr), indicating that with the magnetic field there will be a net increase in angular magnification by the collection lens.  Third, the new angle distribution will break symmetry around the normal emission since the middle red ray is no longer centered between the lower and upper red rays, resulting in uneven angular dispersion across the detector. Fourth, that the new electron rays, projected backwards from the field-free region, converge at a new virtual source point that is displaced from the original source by a displacement vector \vecd\, of magnitude $\sim$ few 10s of \um\ laterally and slightly off the sample surface. To maintain the optimal analyzer lens focus, it is therefore necessary to aim the detector towards the virtual source point. The fifth aberration that is quite important arises for electrons that are emitted out of the plane of \figSchematicRef(b); these effects will be discussed later.

Although the angular imaging properties are affected by the field, \figSchematicRef\ shows that the resulting aberrations are comparable to intrinsic aberrations of the electron lens that arise due to small misalignment of the photon beam to the analyzer axis, as well as residual stray electric and magnetic fields that are commonly found in ARPES vacuum chambers.  The remainder of this paper will describe these aberrations in more detail using simulations of a practical device that approximates the magnetic field conditions in \figSchematicRef. These effects will be confirmed on measurements conducted with a test device.
 
\section{Magnetic Device Design}
\figDevice

Based on the model in \figSchematicRef, we aim to create a magnetic yoke device with a fringing field on the order of 100 mT (or 1000 G), confined to a 2.5 mm $\times$ 100 \um\ $\times$ 100 \um\ volume.  Such a device is shown in \figDeviceRef(a).  

The magnetic field is generated by an H-shaped nickel yoke, magnetized by dual coils driven by currents typically in the range of $\pm 1.0$ A (DC).  We then reduce the current to zero leaving a remanent field in the nickel. This is mostly contained as flux circulating within the yoke body (orange arrows in \figDeviceRef(a)), but a strong localized fringe field will escape at a 100 \um\ gap in the device center.  A sufficiently thin sample is to be mounted over this magnetic gap using conductive epoxy.  The inset shows a typical magnetic device, attached to a 25.4mm diameter sample holder, with two small samples, each 50 \um\ thick, affixed in this fashion and false-colored in green.

The range of remanent fields can be estimated using a simple magnetic circuit model of the H magnet with a vacuum gap\cite{erickson_2020}.  Under this model, a yoke with two 60-turn coils driven by 1A of current, a 28 mm flux path length, and a 100 \um\ gap can produce a maximum field of 332 mT between the poles. This assumes a relative magnetic permeability of 110 (110-600 is typical for nickel).  Based on a retentivity in the range 31-68\% \cite{edwards_magnetic_1927}, this implies a maximum field between the cores in the range 100-225 mT. As we shall see, this leads to maximum useful fringe fields (in remanence) in the range 80-160 mT.

The remanent magnetic field distribution in vacuum was then simulated using SIMION\cite{simion_2}. The device was modeled with two poles having 1.5 mm depth and 1.5 mm width separated by 100 \um\ gap and applying a magnetic scaling factor of 1 mm.  For visualization of the resulting field distributions, we employed a 1 \um\ mesh in SIMION calculations. 

The magnetic field distribution \Bxyz\ may be characterized by a single value defined at the device origin (as defined in \figDeviceRef(b)). The magnetic configurations are labelled by symbols \bup\ and \bdown\ depending on the magnetization direction. For example, a field of \bup=\borigin\ field at the device origin represents, for the given field configuration, the highest in-plane magnetic field that may be achieved as the sample thickness approaches zero.  This corresponds to a  field of \bpoles\ at the center of the gap between the poles.

\figMag

A visualization of the calculated magnetic field distribution is shown in \figMagRef. In \figMagRef(a, b) we show the total magnetic field \absBxyz\ in the (horizontal, vertical) planes ($y=0$, $x=0$), respectively. In the horizontal plane (a), the field, oriented purely along $y$, is shown to be nearly invariant along the gap.  This shows that a sample placed anywhere along the gap will feel the same \vecB.  Furthermore, along the $z$ direction, the field falls off rapidly, with a length scale comparable to the magnetic gap dimension of 100 \um, as desired.  

In the vertical plane shown in \figMagRef(b), we see that the field is confined closely to the magnetic gap region.  We can see from the magnetic flux lines shown (dashed lines) that the fringe fields are curved just outside the yoke surface ($z \gtrsim 0 $).  This implies that away from $y=0$,  a significant out of plane component $B_z$ develops.  The orientation of the local magnetic field therefore varies across the magnetic gap from purely in-plane at $y=0$ to nearly fully out of plane at the edges of the magnetic gap. This field orientation \vecB\  is visualized in \figMagRef(c) as a vector field, where the magnitude of the arrows is proportional to the total field \absB\ and the angle of the arrows gives the field direction.  

 This curvature of the flux lines indicates that samples can be probed with both in-plane and out-of-plane fields, by simply scanning a small probe beam from the center to the edge of the magnetic gap.  The in-plane direction relative to the crystal is chosen by orienting the sample before fixing it to the yoke surface.  Thus, by choosing both the probe position as well as the azimuthal orientation of the crystal during mounting, the full 3-dimensional B-field direction relative to the crystal may be controlled. The uniformity of the field along the gap opening shows that multiple samples can therefore be placed along the gap, e.g. at different crystallographic orientations, in order to systematically probe in-plane anisotropy effects for samples placed side-by-side.  We also observe from \figMagRef(c) that while reversing the in-plane fields probed near $y=0$ requires reversing the field orientation, reversal of the out-of-plane magnetic field may be achieved merely by comparing measurements at the two edges of the yoke gap.
 
\section{Electron Trajectory Simulations}

The simple uniform magnetic field shown in \figSchematicRef (a) does not capture all the effects of the realistic magnetic device and non-uniform fields in Figs. \ref{fig:figDevice} and \ref{fig:figMag}, especially with regards to those electron trajectories for which the free electron velocity \ve\ is not perpendicular to \vecB.  While the latter experience a rigid shift of the angular distribution around the invariant field direction \vecB, other trajectories will exhibit additional helical distortions due to fringe field curvature where $y \ne 0$ and in general for electrons with elevation angle $\beta\ne 0$.  

In order to understand the expected emission patterns, we calculated electron trajectories using the program SIMION\cite{simion_2}, with a geometry comprised of two poles separated by a 100 \um\ gap and modeled within a 10 \um\ cubic mesh for the full 35mm of propagation to the detector input aperture in a magnetic field calculated as described in the previous section.  We assumed for simplicity that the sample had the same magnetic susceptibility as the vacuum.   

For the simulations presented next, we used values \bupdown$=\pm$\borigin\  (=\boriginG), the characteristic field at the gap center $(x, y, z)=(0,0,0)$, and  electron trajectories originated from probe position $(x, y, z)$=(0, 0, 50 \um). This represents emission from a hypothetical surface of the sample of thickness 50 \um.  For this thickness, the in-plane field $B_y$ at the probe position is $\pm$\bsample.  The electron trajectories were then simulated as a function of kinetic energy, takeoff angle, and originating position $(x,y)$, providing a comprehensive analysis of their behavior in this field configuration.

\figDistort

\figDistortRef\ illustrates the resulting angular distributions with and without magnetic field  for electrons with energy of 150 eV.   In \figDistortRef (a), we present simulations for electrons emitted into a 3\degr $\times$ 3\degr grid, while in \figMagRef (b), we simulate the K positions for a hexagonal 2D material with lattice constant 2.46 \r{A} (appropriate for graphene or graphite). The trajectories are then simulated with devices configured as described above. The left panels show that as expected, the angular distribution for $B=0$ are substantially shifted in the polar angle direction ($\theta$) towards (positive, negative) values with magnetic fields \bupdown, resp.  Furthermore, there are significant local aberrations.  

To get a feel for the field-induced imaging aberrations, in the right panels we overlay the cases $B=0$ and \bup; such a comparison would be obtained experimentally by rotating the device polar angle $\theta$ after magnetizing to recenter the emission pattern onto the detector.

For electrons with elevation angle $\beta=0$, the overall effect is a rigid shift of the trajectories along polar angle $\theta$, because these electrons' velocities are purely perpendicular to the field (as in \figSchematicRef).  The angular dispersion of the electrons along $\theta$ also increases, with the largest rate of dispersion observed for the trajectories at larger $|\theta|$.  For electrons emitted with  non-zero elevation angle ($\beta \ne 0$), the electron paths take on a spiral character as they follow the magnetic field lines.  This causes local rotations in the emission patterns. This leads to a reduction of the angular dispersion along the $\beta$ direction,  that has greater effect as both $\beta$ and $\theta$ increase in magnitude.  

The smallest aberrations occur around the trajectories emitted close to the surface normal, showing that electrons emitted near normal emission will hardly be distorted by the magnetic field except for rigid shift in polar angle.  On the other hand, the greatest aberrations are on the high polar angle side, especially at large $\beta$.  The overall affect is the high polar angle side is stretched horizontally, and compressed vertically, with strong local rotations in the emission patterns at large $\beta$. 

For the above simulations, we restricted the magnetic field at the probe site to be in the in-plane  direction (\vecB=(0,$B_y$,0)) by choosing the origin of the electron emission (the probe beam) to be at $y=0$. However as discussed earlier, significant out-of-plane magnetic field can be found as the point of emission approaches the edges of the magnetic device gap.  In \figMagYZRef(a), the black dots and lines show the angular distribution of the graphene K points in the \bup\ configuration discussed above for $y=0$.  The dots with red and blue colors show how the K points redistribute as the probe location $y$ varies from -120 \um\ to 80 \um, corresponding to emission source varying from 70 \um\ below to 30 \um\ above the gap edges.  The mixing  of out-of-plane field away from $y=0$ introduces a clearly visible helical distortion, as well as an overall shift of the pattern back towards $\theta=0$  due to the reduction of overall field experienced by the electrons.

This shift towards $\theta=0$ is expected because of the strong confinement of the magnetic fringe fields to within the gap region, illustrated in \figMagRef(b).  But the observed shift and helical distortion persist far from the magnetic gap edges at $y=\pm 50$\um\ and remain significant to at least $|y|>500$\um.  This is due to remaining weak fringing fields spreading outwards from the gap region that persist all the way to the detector aperture.  Although electrons emitted from positions outside the gap do not feel strong local fields at the point of emission, nevertheless the residual weak fields are still impactful over the long distance to the detector.

We can therefore consider the impact of field on the trajectories to occur in two regimes. First, a ``near-field" interaction that occurs within a few gap spacings of the yoke surface, and mostly affects electrons emitted within the yoke gap. This near-field interaction is approximately visualized in the simplified model of \figSchematicRef (b). Secondly, there is a ``far-field" interaction covering the remaining distance to the detector, that affects electrons emitted both near and far from the yoke gap nearly equally.  This interaction is absent in the simplified model. Although the residual field in the realistic model is very weak in the far-field region, it can still impose a significant angular shift upon the electron trajectories, comparable in magnitude to the angular shift caused in the near-field region.

Aside from the predicted angular shifts, our model also implies a shift in the virtual source position \vecd, which can be thought of as the apparent source position seen by the detector. While \vecd\ is very small in the near-field interaction, (smaller than the gap size and comparable to the 10 \um\ probe size employed), our calculations show a much larger far-field contribution to \vecd.  In \figFocusRef\ we show simulations of \dpar, the  projection of \vecd\ onto the surface plane, for the same \bup\ magnetic configuration discussed above.  Just as we argued for the angular shift, the virtual source position is shifted by almost the same amount -- about 0.80 mm -- regardless of whether the probe is located over the high-field region in the gap.  

Such a large shift in the virtual source position is significant with respect to the electron detector lens column, and would result in significant additional aberration of the trajectories inside the lens which is designed for a source to be located paraxially to the lens column.  Therefore, in order to compensate the effect of magnetic field, it is not only necessary to adjust the polar angle of the sample to return the bands to the detector in angle-imaging mode, but also to laterally shift the detector relative to the probe position so that the virtual source is re-centered in real-space imaging mode.  This can be accomplished by mutually moving the photon probe beam and sample by an amount $-$\dpar\ that can be easily and accurately measured  by the photoelectron detector.  (In principle a compensation of the perpendicular component of \vecd\ is also needed, but for a detector of reasonable depth-of-focus (typically a few mm) this correction is not important.)

\figMagYZ
\figFocus

\section{Experimental Tests}

We conducted experiments using epitaxial graphene grown on 4H-SiC(0001) substrates (Si-face), which were thinned down to 50 µm by lapping of the obverse C-face. The growth was performed on the Si-face using the Ar assisted sublimation technique \cite{Emtsev2009}, which included hydrogen etching by annealing in 1000 mbar hydrogen with a flow rate of 0.5 standard liter per minute (slm) at 1475 °C for 15 min, followed by annealing in 1000 mbar Ar with a flow rate of 0.1 slm at 1475 °C for 15 min. The process was carried out in a dedicated, custom build growth furnace \cite{ostler_2010}.  Two such samples were mounted to our device using conductive epoxy, see \figDeviceRef(a, inset).  The sample heights ($50\pm 1$\um) above the yoke surface were verified by scanning the focal plane of a high-power optical microscope through the sample surface.  Although we could have achieved higher fields by using thinner samples, the chosen samples have the advantage to be highly uniform spatially, so that any angular and source point changes observed could be attributed to the magnetic field device and not sample inhomogeneity.  Experiments were conducted in the \uARPES\ chamber at the MAESTRO beamline at the Advanced Light Source.  Photoelectrons were detected using a Scienta R4000 analyzer outfitted with an imaging channel plate detector and internal deflection plates so that the elevation angle $\beta$ of the photoelectrons could be probed without tilting the magnetic field out of the scattering plane, see \figDeviceRef(b).  

The benefit of using wafer-grown epitaxial graphene was three-fold.  First, graphene provides  useful control data because the bands are sharp and well-localized in angle space, and furthermore, we don't expect to see a strong influence of magnetic fields on the physical properties (for our accessible field range at least).  Thus graphene is ideal as a control material to characterize the magnetic device performance.  Second, the deflection of the graphene electrons emitted at $y=0$, can straightforwardly serve as a magnetometer to measure the device field. This is achieved by comparing their deflections in \bupdown\ to simulations as mentioned above.  And thirdly, its uniformity is critical to measure the general dependence of the angular deflections with arbitrary probe position, in order to verify the above simulations.

The nickel magnetic yoke was fabricated from 1.5mm thick nickel 200 alloy sheet metal (McMaster \#9197K511) and cut to dimensions shown in \figDeviceRef(a) using a wire electrical discharge machine.   Two 60-turn windings were installed on the device's arms. DC current of $\pm$ 1A was applied so that the induced fluxes are applied in parallel.  Current was reduced to zero leaving a strong remanent up or down field, \bupdown, respectively.  MagnetoARPES measurements were conducted with nominal sample temperature $<$ 10 K. The actual temperature is likely to be modestly higher due to low thermal conductivity of the device.

Now we have to contend with the fact that we do not independently know the magnetic field at the sample surface, apart from rough estimates based on material properties and known driving current, because \bupdown\ is impractical to measure due to the small device dimensions.  Our strategy is therefore to use graphene photoemission itself as a magnetometer. By imposing a remanent field and measuring the electron deflection for any one convenient trajectory (electrons at one of the  Dirac crossings at the K point in our case), we can adjust the magnetic field parameter in SIMION until the predicted deflection matches the experimentally observed deflection.   Thus, we need to fit the SIMION model to the data, but this requires only one experimental scaling parameter. The remaining important parameters (electron kinetic energy, gap size, sample thickness and position of the probe beam) are all experimentally known.

Experimental measurements for electrons emitted from the central plane ($y=0$) are reported in \figMagRef(c). In the left panel we show a composite of the angular emission of the graphene K points (acquired in a 15 meV window around the Dirac energy) for three different field magnetic field configurations: \bdown, $B=0$, and \bup.  

Based on comparison of the shift in the bands at K to the simulations, we determined the magnetic field  to be \bupdown=$ \pm$ \borigin\ at the yoke surface, which was thus the reason this value was chosen for the simulations presented in the previous section.  As a result, the experimental angular patterns in \figMagRef(c) are in all aspects in excellent agreement with the predictions in \figMagRef(b).  This agreement is not only in the overall shift of the patterns along $\theta$, but also with respect to the other fine details, including the increased angular spread in the $\theta$ direction, and the reduced angular spread in the $\beta$ directions, which is especially stronger at large values of $|\theta|$.  This shows that the single simulation parameter (central field strength) is able to accurately predict the observed angular deflections as a function of source emission angles $(\theta, \beta)$ for a given kinetic energy and field configuration. 

We now show that this agreement is equally good with respect to probe position $y$.  We  scanned the probe beam along $y$ in the \bup\ configuration, using the same values of $y$ that were simulated in \figMagYZRef(a).  We recorded both the angular emission, using our detector lens column's ``angle-mode" as well as virtual source position, by recording an image of the source using the lens column's ``transmission-" or real-space mode.  

The real-space images collected  are interpreted as images of the virtual source point \dpar\ projected onto the sample plane.  For \vecB=0, this corresponds to the actual source point, which is adjusted in our initial alignment process to appear at position $x=0$ on our detector.  (We are not particularly sensitive to the out-of-plane virtual source position \dperp\ because our electron detector's lens column has a fairly large depth of focus).  After creating the \bup\ field distribution, measurements of \dpar\ as a function of probe position $y$ are determined directly and plotted in \figFocusRef\ and are in excellent agreement with the simulations with respect to both in the the large overall shift ($\sim 0.8$ mm), as well as the smaller relative changes ($\sim 0.1$ mm) in the vicinity of the yoke gap. 

We then adjusted the probe and sample positions to recenter the virtual source onto the detector axis before collecting angle-mode data as a function of probe position $y$. The resulting angular distribution maps of the graphene K points comprise a stack of images that were summed to create a composite map, shown in \figMagYZRef(b). The loci of the rightmost three K points were captured in the resulting map, and these are in excellent agreement with respect to the simulations.  The rightmost K point, at $B=0$ and $\beta=0$, takes a parabolic trajectory in angle space as a function of $y$, having the appearance of a ``lobster claw".  Similar claws at $|\beta|\sim$ 15\degr are also evident. These out-of plane claws are observed to be rotated and pinched somewhat closed compared to the right-most K point, which are nuances that are fully captured by the simulations.

\section{Discussion}

The experimental outcomes are in excellent agreement with the simulations and establish magnetoARPES as a viable technique to study materials in a low but significant magnetic field regime. Not only in-plane but also out-of-plane magnetic fields can be applied, and their effects separately measured through use of focussed x-ray probes in modern microARPES and nanoARPES setups.  The excellent agreement between simulations and observed changes in the angular distribution, as well as observed shift in the virtual source side, indicates that the electrons themselves (in non-interacting samples) in combination with readily available commercial simulation software, may serve to calibrate the magnetic fields produced.  For the $\sim 100$ mT range we explored, these fields should be more than sufficient to observe effects associated with chiral currents \cite{behrends_visualizing_2016}, for example.

As a guide to future experiments, we would like to discuss some important  experimental considerations: 

(1) \emph{small x-ray probe beams}.  The small magnetic gap employed has enabled us to concentrate the field to a small volume, thus minimizing aberrations. But the resulting field distributions are non-uniform,  so a small vertical probe size is necessary to avoid smearing of the beam in angle space.  We recommend for the 100 \um\ gap device shown here, that the probe beam should be no larger than 10 \um\ for in-plane studies at the gap center, and preferably down to 1 \um\ or so if out-of-plane fields are to be probed.

(2) \emph{thin samples}. For maximizing the magnetic fields, it is essential that the samples must be thin enough to ensure their surfaces are well within in the near-field region.  In our case we used thin (50 \um) SiC wafers, which achieved a surface field strength about 60\% of the asymptotic limit at zero thickness. To achieve the highest fields, we recommend thinning the samples to no more than 10\% of the magnetic gap dimension. For bulk crystals that are post- or tape-cleaved, we recommend pre-slicing the crystals down to  $\sim$50 \um\ thickness to ensure the crystals will be thin enough after cleaving.  For exfoliated few-layer 2D materials, which are commonly prepared on typically commercial silicon wafers,  we have experimented successfully and obtained magnetoARPES data for graphene deposited onto 10 \um\ thick Si substrates with native oxide that are commercially available (Virginia semiconductor).

(3) \emph{Distinguishing intrinsic magnetic effects}. The observed aberrations in graphene are purely attributable to extrinsic effects of the magnetic field; direct physical effects on the graphene bands were not sought nor observed in the present study, which should serve as a useful guide to separating extrinsic field effects from instrinsic ones that are of ultimate interest.  Because of the notable local dilations and helical distortions of the maps, intrinsic changes in Fermi surface shape, for example, are relatively difficult to identify.  Another effect difficult to distinguish would be the change of electron lifetime, which would appear as a broadening of the bands that could be due to extrinsic angular smearing of the bands. Thus, any effect that could be attributed to changes in momentum distribution of the bands needs to be treated very carefully.   But other possible changes associated with field, such as energy splittings, the appearance/disappearance of new states associated with altered symmetry, or  shifts of spectral weight in energy would be unambiguously due to intrinsic effects.

(4) \emph{sample temperature}.  We employed nickel as the material for the yoke, but it should be noted that nickel's magnetic permeability is temperature-dependent\cite{harrison_variation_1904}, so the device may have different properties at different temperatures. 

(5) \emph{sample magnetic susceptibility}.  We have ignored the sample susceptibility itself, but this will not be possible for ferromagnetic or diamagnetic materials which may interact with and distort the applied field. In such cases, more detailed magnetostatic models of the sample+device system must be considered to understand the field distribution.

(6) \emph{maximium achievable fields}. We have chosen a relatively high photon energy ($>$ 150 eV) in this study, not because it is crucial to the experiment, but mainly in order that the graphene K points subtend a small enough solid angle matched to our detector's acceptance. Typical ARPES studies employ photon energies down to 20 eV.  Clearly higher kinetic energies favor higher obtainable magnetic fields, but the limitation depends partly on kinetic energy but also on geometric factors such as maximum polar angle achievable.  

Significant gains in field strength could be achieved by miniaturizing the device further.  Employing a 10 \um\ gap and using a 100-1000 nm probe size for example could yield a higher field, but it would require overcoming several difficulties, including thinning the sample to the 1 micron range,  and fabricating the device. We estimate 5 times larger field could be obtained, but this may require using a yoke material with higher permeability, or by conducting the experiments with actively applied current.  We have not explored operating the device with actively applied currents, because of the possible additional stray field effects; this will be explored in a future study.

\section{Acknowledgements}
    We thank Prof. Na Hyun Jo (U. Michigan) for useful discussions.  This material is based upon work at the QSA, supported by the U.S. Department of Energy, Office of Science, National Quantum Information Science Research Centers.  This research used resources of the Advanced Light Source, which is a DOE Office of Science User Facility under contract no. DE-AC02-05CH11231. This research was supported by Basic Science Research Program through the National Research Foundation of Korea(NRF) funded by the Ministry of Education(2022R1A6A3A03069247)

\section{Author contributions}
     S.R. and E.R. conceived the presented idea. Simulations, device design, assembly, and testing were conducted by G.R., S.R.  ARPES measurements were conducted by S.R. and E.R. with assistance of C.M.J. and A.B.  Epitaxial graphene samples were produced by P.R. and T.S.
     
\section{Competing financial interests}
    The authors declare no competing financial interests.

\section{Additional information}
    Correspondence and requests for materials should be addressed to S.H. Ryu (sryu@lbl.gov) or E. Rotenberg (erotenberg@lbl.gov).

\newpage

\section{References and notes}
\normalem  
\bibliography{Reference} 

\end{document}